\documentclass[aip,jcp,reprint]{revtex4-1}

\usepackage{graphicx}
\usepackage{dcolumn}
\usepackage{bm}
\usepackage{amssymb}
\usepackage{amsmath}
\usepackage{color}

\begin{document}

\title{Evolutionary strategy for inverse charge measurements of dielectric particles}

\author{Xikai Jiang}
\thanks{These two authors contributed equally}
\affiliation{Institute for Molecular Engineering, University of Chicago, Chicago, Illinois 60637, USA}

\author{Jiyuan Li}
\thanks{These two authors contributed equally}
\affiliation{Institute for Molecular Engineering, University of Chicago, Chicago, Illinois 60637, USA}

\author{Victor Lee}
\affiliation{James Franck Institute and Department of Physics, University of Chicago, Chicago, Illinois 60637, USA}

\author{Heinrich M. Jaeger}
\affiliation{James Franck Institute and Department of Physics, University of Chicago, Chicago, Illinois 60637, USA}

\author{Olle G. Heinonen}
\affiliation{Materials Science Division, Argonne National Laboratory, Lemont, Illinois 60439, USA}
\affiliation{Northwestern-Argonne Institute for Science and Engineering, Evanston, Illinois 60208, USA}

\author{Juan J. de Pablo}
\email[Email: ]{depablo@uchicago.edu}
\affiliation{Institute for Molecular Engineering, University of Chicago, Chicago, Illinois 60637, USA}
\affiliation{Materials Science Division, Argonne National Laboratory, Lemont, Illinois 60439, USA}

\date{\today}

\begin{abstract}

We report a computational strategy to obtain the charges of individual dielectric particles from experimental observation of their interactions as a function of time.
This strategy uses evolutionary optimization to minimize the difference between trajectories extracted from experiment and simulated trajectories based on many-particle force fields.
The force fields include both Coulombic interactions and dielectric
polarization effects that arise due to particle-particle charge mismatch and particle-environment
dielectric contrast. The strategy was applied to systems of free falling charged granular particles in vacuum, where electrostatic interactions are the only driving forces that influence the particles' motion.
We show that when the particles' initial positions and velocities are known, the optimizer requires only an initial and final particle configuration of a short trajectory in order to accurately infer the particles' charges;
when the initial velocities are unknown and only the initial positions are given,
the optimizer can learn from multiple frames along the trajectory to determine the particles' initial velocities and charges.
While the results presented here offer a proof-of-concept demonstration of the proposed ideas, the proposed strategy could be extended to more complex systems of electrostatically charged granular matter.

\end{abstract}

\pacs{Valid PACS appear here}


\maketitle 

\section{Introduction}
Electrostatically charged granular particles are important in a wide variety of applications, ranging from particulate matter pollution to industrial handling of pharmaceutical products, food grains, and inks for printing and additive manufacturing, to name a few.\cite{dammer200411,Bazant2010}
Granular dielectric particles often acquire charge through tribocharging or contact electrification\cite{henning2000,grzybowski2003electrostatic,Schein2007,kumar2008,sankaran2009,pahtz2010,kumar2014spreading,cai2015};
the charges they carry can significantly affect their dynamics and their interactions with the surrounding environment.
In order to better understand how such charged, polarizable particles interact, it is therefore of fundamental importance to make detailed measurements of their actual charge\cite{hyde2014}.
Recently developed experimental techniques have made attempts to determine the charges of individual particles in a vacuum environment using free-fall videography\cite{royer2009,jaeger2013charge,waitukaitis2014prl}. In those experiments, particles falling under the influence of gravity were filmed as they interacted in a vacuum tube.
In one experiment, by accelerating charged particles in a horizontal electrical field and analyzing approximately $\sim10^4$ trajectories\cite{jaeger2013charge}, the average particle charges were estimated by relying on the relationship between acceleration, mass, and charge.
In another experiment, by identifying the relative positions of two particles and fitting Kepler-like orbits\cite{lee2015nphys} to their motion, it was possible to determine their charges.
The interactions that arise amongst polarizable particles are inherently many-body, and it is therefore essential that new approaches be developed that are capable of taking such effects into account. In this work, an approach is proposed that is capable of simultaneously measuring the charges of many individual particles from a {\it single} set of trajectories (i.e., the trajectories of the particles from a single experiment, as opposed to an ensemble of trajectories from many different experiments). The approach relies on two advances: (1) the availability of new numerical algorithms and new analytical expressions capable of describing polarizability effects on interacting particles \cite{holm2010,bertiPRE,polarizationMonica,barrosPRL,barrosJCP,jiang2016on,xuPRE,Freed:2014fz,qin2016image1,qin2016image2}, and (2) the availability of modern evolutionary computation strategies \cite{fogel1966,hansen2001,jin2005,YANG2008} that enable direct interpretation of experimental data from numerical computer experiments.

\section{Methods}

\subsection{Inverse problem}

\begin{figure*}
\centering
\includegraphics[width=0.6\textwidth]{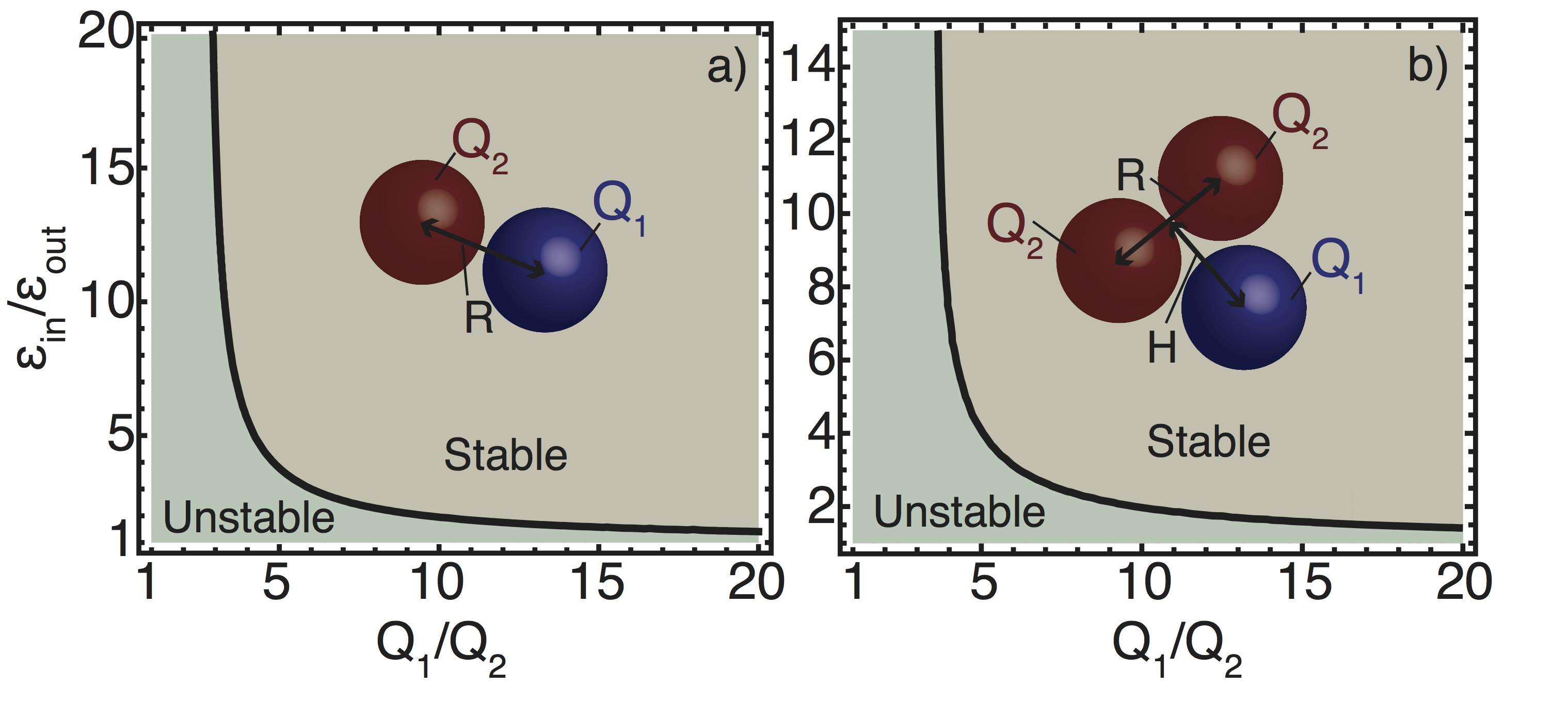}
\caption{Stability diagram for dimer and trimers.
Clusters of like-charged particles in close contacts are stabilized by surface charge polarization.
The parameter regimes in which the close-contact particle aggregates
are stabilized are highlighted with colored shades.
The boundaries between different regimes are identified by
computing the gradient of energy with respect to particle displacements. Notice that all particles here are positively charged and the different charge amount is labeled by red and blue color.}
\label{fig1}
\end{figure*}

To computationally determine charges on granular particles from a given single set of target trajectories assembled on a time-sequence of $N_f$ frames, we adopt an evolutionary optimization technique that seeks to minimize a fitness function. Here that function $f$ is defined as the deviation between trial trajectories generated in each optimization step and the target set of experimental or computational trajectories:
\begin{equation} \label{fitness_f}
f=\frac{1}{N_f} \sum_{k=1}^{N_f}{\left( \frac{1}{{N_p}} \sum_{i=1}^{N_p} {\vert \boldsymbol{r}_{i,\mathrm{trial}}^{(k)} - \boldsymbol{r}_{i,\mathrm{target}}^{(k)} \vert} \right) },
\end{equation}
where $N_f$ excludes the initial configuration, 
$N_p$ is the number of granular particles,
$\boldsymbol{r}_{i,\mathrm{trial}}^{(k)}$ and $\boldsymbol{r}_{i,\mathrm{target}}^{(k)}$ are the positions of the $i$-th particle at $k$-th frame
in the trial and the target trajectories, respectively. If the masses of the particles are known, the trial trajectories can be obtained using simulations with a suitable force field, i.e. in this case the electrostatic interactions, which include pair-wise Coulombic forces and many-body dielectric polarization contributions. The bare Coulombic interaction can be attractive or repulsive, depending on the sign of the charges. Polarization effects, however, are purely attractive when the internal dielectric permittivity of materials is greater than that of the medium, and are purely repulsive in the opposite case \cite{qin2016image2}. This polarization-induced attraction is summarized in Fig.~1 for two and three particles of equal-sign charge. The figure shows the conditions, $\epsilon_{in} / \epsilon_{out}$ and $Q_1/Q_2$, under which two and three particles will form stable (attractive) aggregates. The boundary between the stable and unstable states is calculated using a recently developed analytical, perturbative theory \cite{qin2016image2}(see Sec. II B below).
It is clear that dielectric polarization can strongly influence the nature of interactions between charged dielectric objects, particularly when sharp dielectric discontinuities are involved. In this work, we use the recently proposed analytical formalism (image method) to calculate electrostatic interactions between polarizable granular dielectric particles, and we have implemented the resulting electrostatic force field into LAMMPS (http://lammps.sandia.gov)\cite{PLIMPTON1995JCP} to simulate trajectories of the particles.

A Covariance Matrix Adaption Evolution Strategy (CMA-ES) is adopted, and we rely on the open source library libcmaes (https://github.com/beniz/libcmaes)\cite{benazera2014libcmaes} to extract the charges
through an iterative optimization process.
We address the inverse problem under two scenarios. In the first, the initial velocities of the particles are known, but their charges are not. The search variables are therefore the $N_p$ charges.
In the second scenario, both the initial velocities and charges of the particles are
unknown, so there are in total $4 N_p$ search variables ($N_p$ charges and $3 N_p$ velocities in 3-dimensional (3D) space).
As our results demonstrate, the proposed strategy is able to determine the charges of the particles under both scenarios.
It is difficult to know how many local optima may be encountered in the fitness landscape. In our simulations, we find that the optimizer can be trapped in one of several optima during a series of consecutive optimizations for the test cases in Sec. III B and C, which suggests that multiple optima are generally accessible. To help the optimizer escape a local optimum, and move towards the global optimum, we restart the optimization and rescale all search variables as shown in Sec. III B and C. Mathematically, the global optimum is found when the fitness function decays to zero; numerically, the global optimum is found when the fitness function is smaller than a tolerance. The value of the tolerance depends on the underlying errors in the experimental trajectories as well as on the numerical approximations involved in the particle simulations.

\subsection{Image method}
Image method is an analytical method capable of describing polarizability effects on interacting particles. We consider $N$ spherical particles, with radius $a$ and dielectric permittivity $\epsilon_{in}$, embedded in a continuum with dielectric permittivity $\epsilon_{out}$. In principle, our approach
may be used for polydispersed particles with different $\epsilon_{in}$ and $a$, but we will limit the discussion in this work
to monodispersed systems of equally sized spheres. The $i$--th particle carries a point charge
$Q_i(\mathbf{x}_i)=z_ie$ at the center, where $z_i$ is the valence and $e$ is the elementary charge, which implies a homogeneous free surface charge density on the particle. Note, however, that in future work it should also be possible to pursue the proposed inverse calculations using numerical methods - such as those proposed in our recent work - that take inhomogeneous free charge distributions into account\cite{jiang2016on}. In this work, we use the numerical method to calculate the induced surface charges on particles and found that the induced surface charges are inhomogeneous and their distribution changes as particles move along the trajectory as shown in Fig.~\ref{fig5}.

The main interactions between the particles are the usual pairwise Coulombic interactions.
However, when the particles are in close proximity, they induce surface charges,
that give rise to additional interactions.
We recently developed a systematic multiple-scattering formalism~\cite{qin2016image1,qin2016image2}
to describe this polarization interaction.
In this formalism, the polarization energy is grouped in terms according to the number of interacting particles.
The lowest-order of the polarization energy, i.e., the three-body terms $E_3$, is contributed to by $3$ particles.
The higher-order terms $E_4$, $E_5$ involve four-body, five-body interactions, etc.;
the two-body terms are reserved for the normal pairwise Coulombic interaction.
Symbolically, the total electrostatic energy, $E_E$, for an ensemble of dielectric spheres may then be written as
$E_E = E_2 + E_3 + E_4 + \cdots$,
where each term in such an expansion involves a summation over all possible two-body, three-body, four-body, and so on, permutations.
The key point to note about this multi-body expansion for $E_E$ is that all interaction terms
only depend on the particle positions. The references to surface charges are avoided by
replacing the induced charges by the gradient of the electrostatic potential,
therefore the degrees of freedom are greatly reduced.
Furthermore, the forces on the particles can be computed via differentiation with respect to the particle positions,
which enables $N$--body particle simulations.
Three terms are preserved for the electrostatic multibody potential in the particle simulations, which have been shown to be essential in describing particle interactions in the presence of polarization\cite{qin2016image1,qin2016image2,jiang2016on}.

\subsection{CMA-ES}
CMA-ES is one type of evolutionary optimization algorithm\cite{hansen2016cma} that does not require derivative information of the fitness function,
so it enables minimizing a broad range of fitness functions that have no analytical forms.
In general, the idea of evolutionary optimization is to first generate a sample of random search variables every generation following a Gaussian distribution,
then select the best search variables that produce the most optimized value of the fitness function.
This process is then iterated until the fitness function is within a target convergence criterion.
In CMA-ES, the mean and covariance matrix of the search variables as well as the step size are updated every generation to achieve fast and successful optimization.
CMA-ES has found many applications in various materials design problems\cite{depablo2013,depablo2014,Miskin2016,depablo2016,katz2017,nealey2017}.

\subsection{Particle simulation}
We consider spherical granular particles in a 3D vacuum environment, where only electrostatic interactions are the driving force for their motions.
Recent experiments\cite{lee2015nphys} in a similar setting have observed striking phenomena of aggregation and motion of charged granular particles,
from which we adopt the particles' parameters for our simulations.
Specifically, the particles are monodisperse and have relative dielectric constants of 15,
diameters of 260 $\mathrm{\mu m}$,
and mass densities of 3800 $\mathrm{kg/m^3}$.
No thermal fluctuation or Brownian motion of the particles is included; the particles are in vacuum and have diameters of hundreds of microns.
The system may become chaotic when it is evolved over time scales longer than those used in this work. However, a short-time trajectory without chaotic behavior (maximum of 25 ms in this work) is sufficient to successfully extract the charges of particles using our proposed strategy.
We also neglect the particles' rotational motion and only account for their translational motion because particles studied in this work have spherical shapes.
The charges on every particle are assumed to be uniformly distributed and the electrostatic interactions that include both Coulombic interaction and polarization effect are calculated using the aforementioned image charge method\cite{Freed:2014fz,qin2016image1,qin2016image2}.
To simulate the trajectories of granular particles, Newton's equation of motion is integrated by the velocity-Verlet algorithm in LAMMPS with a time step of 1~$\mathrm{\mu s}$.
For the electrostatic interaction, the boundary condition is such that the electrical potential decays to zero at infinity. The particles are simulated in the \textit{NVE} ensemble without periodic boundary conditions.
In test problems, the initial positions of the particles are randomly generated while ensuring there are no overlaps between any two particles.
For the test problems with ten and 30 particles in this work, the target trajectories are generated by simulations with initial velocities all set to zero. For the problem with ten particles, the particles' charges are $\pm 1, \pm 2, \pm 3, \pm 4, and \pm 5$ pC (picoCoulomb), respectively.
After $t=0$, particles start to move under the influence of the electrostatic forces.
For the problems with ten and 30 particles, the dynamic simulations is run for 2000 steps (2~$\mathrm{ms}$) and we sample and store the trajectories with a frequency of 1 frame per 10 steps (10 $\mathrm{\mu s}$).
There are no collisions between any two particles in the trajectories, so tribocharging phenomena are avoided.
The trajectories generated by these simulation are then imported as the target trajectories to the optimization program for inverse calculation of charges on the ten and 30 particles, respectively.

As alluded to earlier, in this work we assume that it is sufficient to assume a uniform charge distributions on the particles' surface, which is equivalent to placing a point charge in the center of the spherical particle according to Gauss' law\cite{qin2016image1,qin2016image2}. In reality, the free charges on particles' surface may be distributed inhomogeneously\cite{Baytekin308}. According to Ref.~\onlinecite{Baytekin308}, immediately after the contact, the surface charge distributions are inhomogeneous in the contact areas; after 2.2 hours, they become uniform. In our experiments, the particles are placed and stabilized in the chamber for more than 3 hours before their free fall begins, such that the surface charge distributions are uniform. During free fall, there are collisions and contacts between particles, and charges are expected to be inhomogeneous in contact areas. Because the contact area is much smaller than the total surface area of each particle, however, the charges can be reasonably assumed to be mostly uniform on the particles' surfaces. As a result, the charge non-uniformity has a minor effect on the particles' trajectories. Ideally, we would include charge non-uniformity in our simulations and estimate charges on individual patches on all particles; this could be done by relying on numerical methods such as those introduced in our earlier work\cite{jiang2016on}, but at greater computational expense. If there are many patches of charges, we conjecture that the number of solutions for the charges that can match experimental trajectories will still be one (with nonzero external electric field), or two (without external electric field, where the signs of the charges will be opposite in the two solutions). As the number of patches increases, the fitness function landscape becomes rougher, and the number of iterations to converge and the associated computational cost increases. Thus, using an efficient electrostatic solver\cite{jiang2016on} would help reduce the computational cost for this problem. However, in view of the lack of systematic experimental data on how the charges are really distributed on the particles during their free fall, we find it difficult to assume a certain pattern of non-uniform surface charge distributions on the particles. By instead assuming a uniform distribution, we are still able to produce simulated trajectories that agree very well with experimental trajectories, serving to validate our assumptions. Although the free surface charges are assumed to be homogeneous, we find that the induced surface charges on particles are inhomogeneous by calculating induced charges using the numerical method, and the distribution of the induced surface charges changes as particles move as shown in Fig.~\ref{fig5}.

\section{Results and discussion}

\subsection{Known initial velocities}

\begin{figure}
\centering
\includegraphics[width=0.3\textwidth]{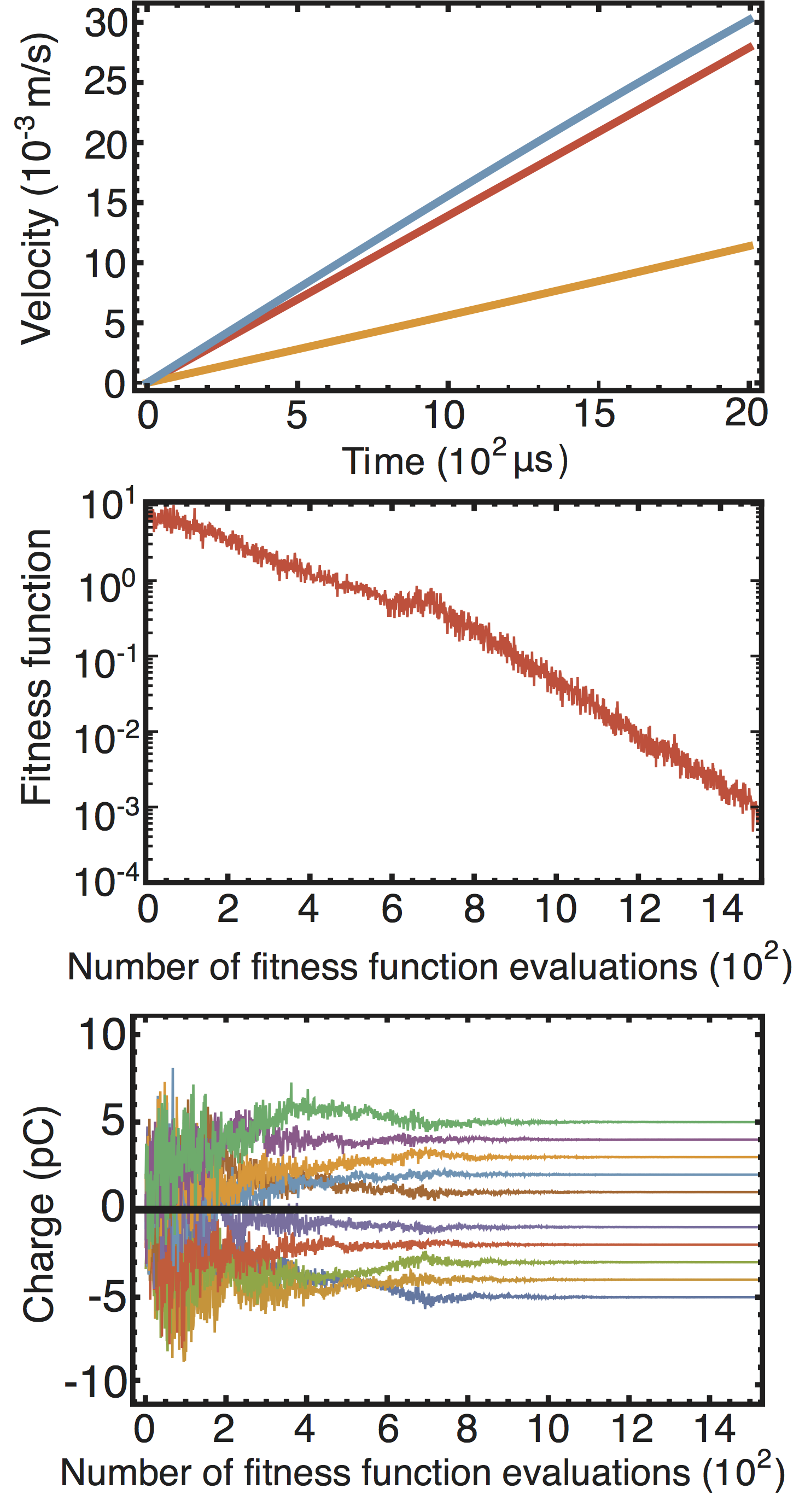}
\caption{Top panel shows the evolution of velocities of three representative particles as a function of time in the target trajectory (solid line) and in the trajectory generated by the simulation using inversely calculated charges (dotted line).
Middle and bottom panels show the evolution of the fitness function and charges of 10 individual particles, respectively, as a function of the number of fitness function evaluations.
Every optimization step contains complete trajectories of ten electrostatically charged granular particles.}
\label{fig2}
\end{figure}

We first study the inverse problem when the charges of the particles are the only unknowns; the known particles' initial positions and velocities are used to start the simulation.
In our test case, we first generate trajectories of ten charges with assigned charges, and we use a single final frame at $t=2$~ms with the final positions of the particles,
so $N_f$ in Eq.~(\ref{fitness_f}) is 1. The true (assigned) charges of the particles are $\pm 1, \pm 2, \pm 3, \pm 4$, and $\pm 5$ pC, respectively, and the aim of this first example is to demonstrate that the evolutionary optimization process can correctly recover those charges from knowledge of the particles' masses, initial positions and velocities.
We use the following three parameters in the CMA-ES evolutionary optimization.
The initial values for charges are set to zero; the initial search step is 2,
and the number of offsprings is 10.
Figure~\ref{fig2}(a) shows the velocities of three representative particles as a function of time throughout the simulated trajectory. One can see that the trajectory generated by the simulation using inversely calculated charges (dotted line) agrees well with the target trajectory (solid line).
Figures~\ref{fig2}(b) and~\ref{fig2}(c) show the evolution of the fitness function and the estimated charges
of ten particles as a function of the number of fitness function evaluations.
The fitness function decreases as the number of evaluations increases,
and it converges to about $2 \times 10^{-2}$ after 110 generations ($\sim1100$ fitness function evaluations).
The deviation between trial and target particle trajectories becomes smaller as the optimization proceeds, and the trial charges on the particles gradually evolve to their target values.
When the fitness function reaches a plateau, the particles' charges stabilize at the correct values of
$\pm 1, \pm 2, \pm 3, \pm 4$, and $\pm 5$ pC, respectively.

To examine how the initial guess for the particle charges affects convergence, we chose four different starting values, i.e., 0 pC, 1 pC, 10 pC, and 100 pC, and examined the evolution of our optimization while keeping all other CMA-ES parameters constant. For an initial guess of 0 pC, 1 pC, and 10 pC, all three optimizations yielded correct estimates of the charges, but the number of generations to reach convergence increased as values of particle charges in the initial guess increased. When the initial guess is 100 pC, the optimizer cannot reach convergence in a single optimization, i.e., the estimated charges at the end of the first optimization are of a different order of magnitude. These findings serve to illustrate that it is critical to choose values for the initial particle's charges that are of the same order of magnitude as the target value, otherwise, rescaling the search variables multiple times is necessary to achieve convergence, as shown in Sec. III B.

In this test case, we also find that using only the Coulombic interactions (i.e. neglect polarization interactions) leads to 10\% error in the inversely calculated charges.
We note that the importance of polarization depends on i) the ratio between the dielectric constant of the particles and that of their surrounding environment, and ii) the charge ratio between interacting particles. The polarization effect becomes more important when the dielectric ratio or charge ratio increases. In the first problem considered here, the dielectric ratio is 15 and the maximum charge ratio is 5. According to Fig.~\ref{fig1}, with this combination of parameters, the polarization effect is not too strong, which is consistent with the 10\% error that is observed when polarization is neglected. However, for particles with dielectric ratios that are larger than 15 and charge ratios that are above 5, neglecting polarization effects leads to errors that are much larger than 10\%; in those cases, polarization effects must be taken into account. Moreover, attractions and adhesions between like-charged particles were observed in Ref.~\onlinecite{lee2015nphys}, where polarization was shown to play a central role.

\subsection{Unknown initial velocities}

\begin{figure}
\centering
\includegraphics[width=0.3\textwidth]{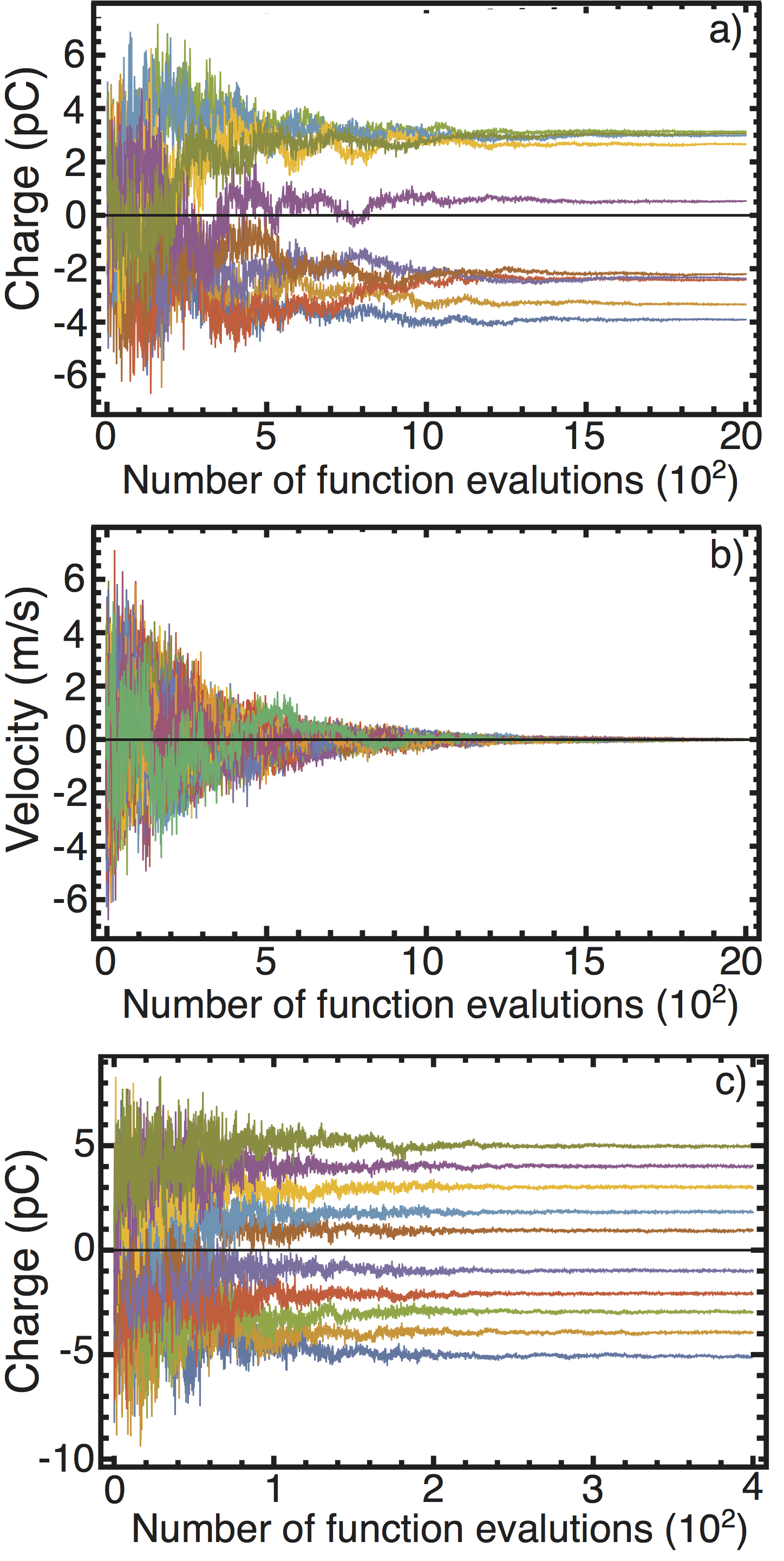}
\caption{Evolution of charges and initial velocities for ten individual particles as a function of the number of fitness function evaluations when the initial velocities are unknown. (a) Evolution of charges in the first optimization; (b) Evolution of initial velocities in the first optimization; (c) Evolution of charges in the second optimization.}
\label{fig3}
\end{figure}

Accurately recording both positions and velocities of granular particles 
using videography is challenging.
In most experiments, only the particles' positions are recorded. It is of course possible to approximate velocities at every frame using a finite difference approximation, but that may lead to a loss of accuracy.
It is therefore of interest to explore the use of \textit{only} information about the positions of the particles for inverse determination of their charge.
We find that, when the particles' initial velocities are unknown, using multiple frames from the particles' trajectories to evaluate the fitness function
can enable such inverse charge calculation.
Specifically, 20 consecutive frames are selected from the initial stage of the simulated trajectories for ten particles (having the same charges as above)
at an interval of 10~$\mathrm{\mu}$s; $N_f$ in Eq.~(\ref{fitness_f}) is set to 20.
The parameters for performing the CMA-ES evolutionary optimization are as follows:
the charges and initial velocities are set to zero for all particles, the initial search step is 2, and the number of offsprings is 10.
Figure~\ref{fig3} shows the evolution of the charges and initial velocities as a function of the number of fitness function evaluations.
As the optimization progresses, the initial velocities are estimated to be close to their true values, as shown in Fig.~\ref{fig3}(b).
Unfortunately, however, there are significant deviations between the estimated charges and their true values, even when the charges stabilize at the end of the optimization process (about 200 generations), see Fig.~\ref{fig3}(a).
The converged value of the fitness function for Figs. 3(a) and 3(b) is 1.28 $\mu m$.

The estimated initial velocities at the end of the optimization process in Fig.~\ref{fig3}(b) are on the order of $10^{-3}$ to $10^{-4}$~m/s, while the charges are on the order of $10^0$~pC. Thus there is a difference of about 3 to 4 orders of magnitude between the numerical values of the estimated charges and the initial velocities. This points to a numerical artifact of the optimization process as originally implemented;
for successful optimization using CMA-ES, it is essential that variables be rescaled in order to ensure that all numerical values of the search parameters are of the same order of magnitude\cite{cmaesHints}. 
Then a second optimization simulation is started by setting initial values for the search parameters to those corresponding to the last step of the previous simulation; in the subsequent simulation, we re-scale the search variables for initial velocities by $10^{-4}$, i.e., if one of the initial velocities in the optimizer is 1, then its value fed to the dynamical simulation is $10^{-4}$.
The results of this second optimization process are shown in Fig.~\ref{fig3}(c).
One can appreciate that the charges evolve rapidly towards their true values after about 30 generations (300 fitness function evaluations). At the end of the optimization process, the charges agree very well with their true values and the converged value of the fitness function for Fig. 3(c) is $3.69\times10^{-4} \mu m$, serving to demonstrate that the inverse calculation process can accurately estimate charges from known trajectories, even if the particles' initial velocities are unknown.

\subsection{Random charges}

\begin{figure}
\centering
\includegraphics[width=0.4\textwidth]{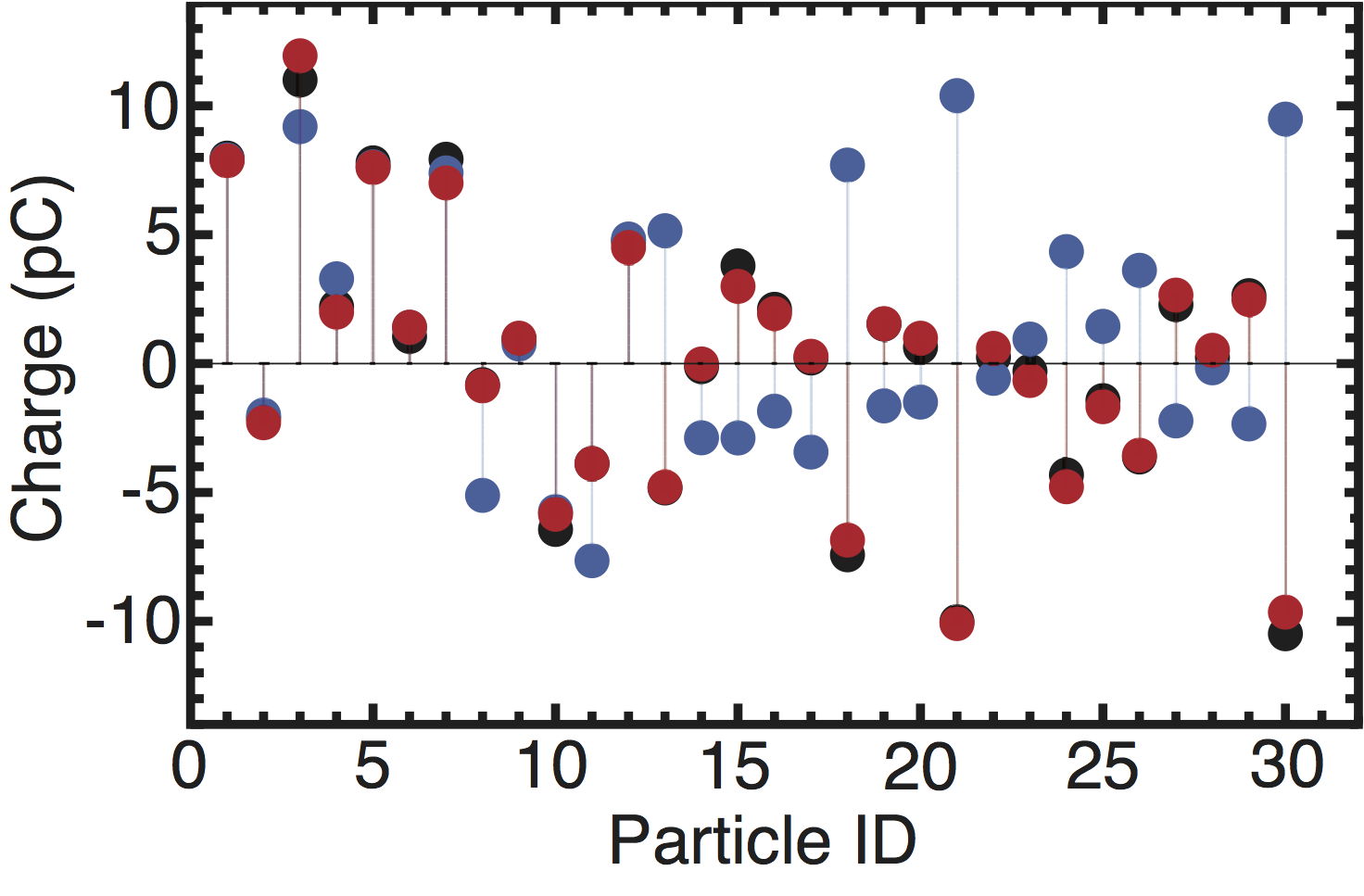}
\caption{Particle charges as a function of particle number. Red, blue, and black dots represent true charges, calculated charges in the absence of the external electric field, and calculated charges in the presence of the external electric field, respectively.}
\label{fig4}
\end{figure}

To further test the robustness and applicability
of our proposed strategy, a third test was performed on a system of 30 particles with randomly assigned charges, drawn from a Gaussian distribution with zero mean and a standard deviation of 5~pC.
Their values are represented by red dots in Fig.~\ref{fig4}.
True trajectories were then generated from simulations using these random charges and zero initial velocities as inputs.
We applied the evolutionary optimization strategy to this problem assuming unknown charges and unknown initial velocities. Multiple optimization simulations were performed consecutively to rescale the search variables properly. As can be seen in Fig.~\ref{fig4}, upon convergence most of the estimated charges (blue dots) do not agree particularly well with their true values (red dots).
In fact for the 13-th to 30-th particles, the signs of the estimated charges are just the opposite of their true values. On the other hand, the absolute values of the estimated charges agree well with the true values.
This happens because of the symmetry of the system, i.e., a trajectory remains the same when the particles' charges all have reverse signs (the mass and shape for all particles are the same).

To resolve the sign problem, an external electric field was applied to break the charge symmetry during the generation of model trajectories. The evolutionary optimization strategy was then used on these new trajectories.
Fig.~\ref{fig4} shows that after optimization with applied field, the charges obtained through the optimization process (black dots) agree very well with the target values.
This result shows that by breaking the charge symmetry by applying an electric field, the evolutionary optimization strategy can correctly recover the charges of the particles.
We have also varied the magnitude of the applied field from 0.1 to 100~V/$\mu$m, and found that all values lead to the correct sign of the charges. The difference between various magnitudes of the applied field is that larger electric fields can generate a trajectory that is different from that without a field within a shorter amount of time.
Since the only good of applying a field is to determine the sign of a charge (and not its magnitude), such differences have no influence on our results.

\subsection{Application to experimental trajectories}

\begin{figure*}
\centering
\includegraphics[width=0.62\textwidth]{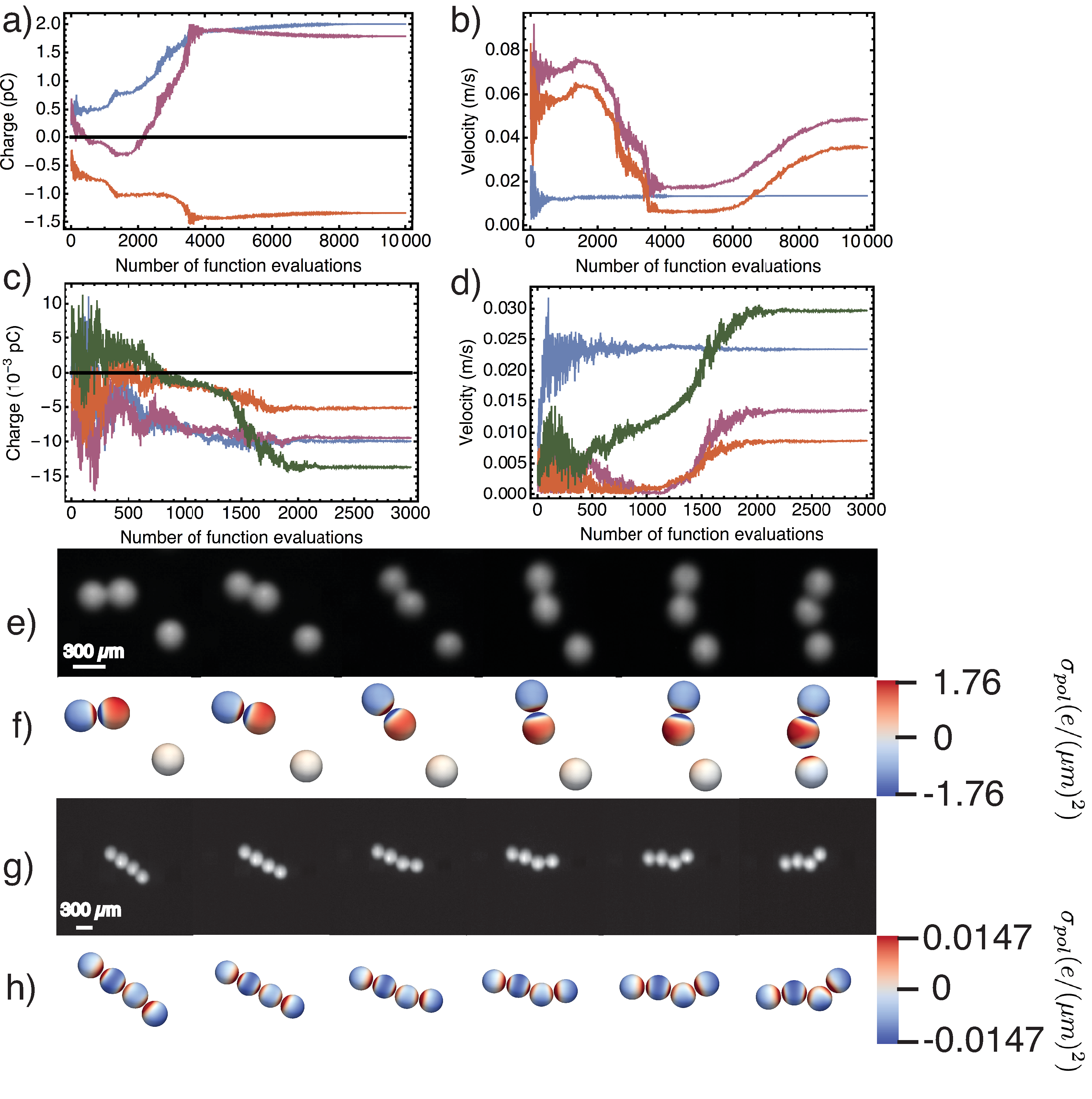}
\caption{(a), (b) The evolution of charges and initial velocities of three individual particles as a function of the number of fitness function evaluations; (c), (d) The evolution of charges and initial velocities of four individual particles as a function of the number of fitness function evaluations; (e) and (f) show snapshots of three particles moving in vacuum environment from experiment (e) and simulations (f), and the time interval between two consecutive snapshots is 5 ms; (g) and (h) show snapshots of four particles moving in vacuum environment from experiment (g) and simulations (h), and the time interval between two consecutive snapshots is 4 ms.}
\label{fig5}
\end{figure*}

Lastly, we apply the evolutionary optimization strategy to experimental data to (i) calculate charges on granular particles in experiments, and (ii) reproduce the experimental trajectories using simulations.
A set of trajectories for three granular particles and another set of trajectories for four granular particles are chosen from the experimental data. The data set, which consists of 25 frames, covers a span of 25~ms. Data were captured from videography; a particle tracking technique was applied to extract the coordinates of all particles. The time interval between consecutive frames is 1~ms.
In the trajectory with three particles, two particles are always in contact with each other, while the third particle moves freely around the other two; in the trajectories with four particles, particles are always in contact with their neighbors. A bond is formed between the two sticking particles by short-range cohesive forces, including van der Waals forces or capillary forces due to absorbed molecular layers\cite{royer2009,salameh2012,salameh2017}.
These short range interactions are strong enough to hold the two particles together without relative translational and rotational momentum between each other. To reproduce this behavior, a rigid bond is implemented in the simulations between the sticking particles. The length of the rigid bond is set to the diameter of one granular particle and is maintained in every simulation step using the SHAKE algorithm\cite{RYCKAERT1977327}.
The charge on each particle is kept constant in particle simulations, because charge transfer between particles is negligible during the short trajectory time of 25~ms\cite{lee2015nphys}.
The coordinates of three and four particles extracted from experiments are then fed as the target trajectory into the optimizer, and the optimization strategy is applied.
Note that in this case the signs of the particles' velocities can be inferred from experimental data, and we constrain the sign of the search variables for initial velocities in the optimization program to increase efficiency.

Figures~\ref{fig5}(a), \ref{fig5}(b), \ref{fig5}(c), and \ref{fig5}(d) show the evolution of charges and initial velocities as the optimization proceeds, for three and four particles, respectively.
The results presented here are from the last optimization simulation in a set of consecutive calculations. In the first few simulations, the fitness function decreases to a plateau at the end of each simulation. However, the fitness function is still large and on the order of $10^2$. Moreover, there is a difference of about three orders of magnitude between some of the search variables.
These consecutive simulations are merely used to properly rescale all search variables; we then feed the final charges and initial velocities to the particle simulation to generate a simulated set of trajectories of three and four particles.
Figures~\ref{fig5}(e) and~\ref{fig5}(f) show six snapshots of three particles from experiments and simulation, respectively.
The induced surface charge density ($\sigma_{pol}$) on every particle is calculated using COPSS (https://bitbucket.org/COPSS/copss-polarization-public)\cite{jiang2016on}.
We obtain excellent agreement between the simulated and experimental trajectories of the particles.
The charges on the three particles are obtained as 1.785, -1.338, and 2.0~pC, respectively (from left to right in the first snapshot in Fig.~\ref{fig5}(f)). They could also be -1.785, 1.338, and -2.0~pC because of the symmetry of the system. The range of calculated charges is consistent with that inferred in previous experimental results\cite{lee2015nphys}; the experimental charge distributions $P(q)$ have tails up to several million electron charges ($10^6 e \approx$ 0.16 pC).
It is also found that the two bound particles carry opposite charge, and the Coulombic attraction force helps bind them together.
For the set of trajectories comprising four dielectric granular particles,
Figs~\ref{fig5}(g) and~\ref{fig5}(h) show six snapshots from experiments and simulations, respectively.
Excellent agreement is again found between both, serving to demonstrate the applicability of our proposed optimization strategy for inverse charge calculations.
The charges on the four particles are obtained as -9.97$\times10^{-3}$, -9.37$\times10^{-3}$, -5.12$\times10^{-3}$, and -1.37$\times10^{-2}$~pC, respectively (from right to left in the first snapshot in Fig.~\ref{fig5}(h)). They could also be of positive sign because of the symmetry of the system. The range of calculated charges is consistent with that inferred from previous experimental measurements\cite{lee2015nphys}. Note that the magnitudes of the charges for four particles are much smaller than those for three particles, and the signs of the four particles' charges are the same, indicating that polarizability is essential for describing the physics of the particles considered here.
The converged values of the fitness function for Figs.~\ref{fig5}(a) and ~\ref{fig5}(c) are 13.66 $\mu m$ and 48.5 $\mu m$, respectively, which are larger than those for Fig.~\ref{fig3}. The converged values in Fig.~\ref{fig3} are relatively small, because i) the target trajectory is generated by simulation and the trajectory data are an exact representation of the simulated particles' motion; ii) there is no error from the numerical simulation results, since the models used in generating the trajectory and in optimizations are identical. The relatively large converged values in Fig.~\ref{fig5} when fitting the experimental data are likely due to errors from experimental measurements and approximations used in the numerical model. The experimental errors are from vibrations of the camera and the particles' position tracking process; note that the resulting errors in the particles' positions are on the order of 10 $\mu m$. The approximations in the numerical model include using rigid bonds to connect sticking particles, and assuming a homogeneous free surface charge density on each particle.
Movies for the above sets of trajectories from experiment and simulation can be downloaded from the Supplementary Material.



\section{Conclusions}
In summary, we have combined an evolutionary optimization strategy CMA-ES with a particle dynamics simulator to obtain the charges on granular polarizable particles based on a given set of experimental trajectories.
The availability of a polarizable force field for electrostatically interacting charged granular particles is central to the particle dynamics simulator; electrostatic polarization and Coulombic interactions can in some cases have opposite signs, and lead to trajectories that are very different from those observed in the absence of polarizability effects.
The proposed strategy was demonstrated in the context of several problems.
In the first problem, the initial position and velocities of all particles were given, and the algorithms were used to infer the particles' charges.
In the second and third problems, both the particles' charges and  initial velocities were unknown, and it was shown that the evolutionary optimization can be used to successfully determine the particles' charges and their initial velocities.
In the fourth problem, the evolutionary optimization strategy was applied to extract the charges from experimentally observed trajectories, and the charges were found to be within the ranges reported in previous experimental measurements from the literature.

The proposed strategy could be extended to more complex systems containing electrostatically charged granular particles.
For example, using a recently developed parallel $O(N)$ numerical solver for electrostatic polarization interactions among arbitrary-shaped particles\cite{jiang2016on}, the evolutionary optimization strategy could be applied to determine charges not only of spherical particles, but also on arbitrarily-shaped particles with uniform or nonuniform surface charge distributions, and including rotational motion.
The proposed strategy could also be used to determine the charges of particles in micro- or nano-fluid environments by coupling the strategy with a recently developed parallel $O(N)$ Stokes' solver for hydrodynamically interacting objects in general geometries\cite{zhao2017on}.
We envision that our proposed strategy could find applications in material property measurements and material designs.

\section*{Supplementary Material}
See supplementary material for the experimental and simulated trajectories of dielectric granular particles.

\section*{Acknowledgements}
The development of the algorithms, codes, and strategy for inverse interpretation of experimental data presented in this work were supported by the Department of Energy, Basic Energy Sciences, Materials Research Division through the Midwest Center for Computational Materials (MICCOM). The development of the COPSS code is also supported by MICCOM. The experimental characterization of polarizable particles and the extraction of particle charge from experimental particle trajectories were supported by the University of Chicago Materials Research Science and Engineering Center, which is funded by the National Science Foundation (NSF) under Award DMR-1420709.
V.L. acknowledges support from NSF DMR-1309611. We gratefully acknowledge the computing resources provided on Blues and Bebop, a high-performance computing cluster operated by the Laboratory Computing Resource Center at Argonne National Laboratory, and the University of Chicago Research Computing Center for support of this work.

\bibliography{references}

\end{document}